\documentclass[12pt]{article}
\usepackage{amsmath,amssymb,amsfonts,amsbsy}
\usepackage{epsfig,subfig}
\usepackage[usenames,dvipsnames]{color}
\usepackage[authoryear,round]{natbib}
\usepackage{algorithm}
\usepackage[letterpaper, left=1truein,right=1truein,top=1truein,bottom=1truein]{geometry}
\usepackage{arydshln}
\usepackage{booktabs} 
\usepackage{multirow}
\usepackage{textcomp}

\def\log{\hbox{log}}

\def\boxit#1{\vbox{\hrule\hbox{\vrule\kern6pt
			\vbox{\kern6pt#1\kern6pt}\kern6pt\vrule}\hrule}}

\def\bse{\begin{eqnarray*}}
	\def\ese{\end{eqnarray*}}
\def\be{\begin{eqnarray}}
\def\ee{\end{eqnarray}}
\def\bq{\begin{equation}}
\def\eq{\end{equation}}
\def\bse{\begin{eqnarray*}}
	\def\ese{\end{eqnarray*}}

\newtheorem{theorem}{Theorem}
\newtheorem{lemma}{Lemma}
\newtheorem{proof}{Proof}

\DeclareMathOperator{\row}{row} 

\title{Non-iterative Joint and Individual Variation Explained}
\author{Qing Feng, Jan Hannig and J.S.Marron}
\author{
	{\sc Qing Feng, Jan Hannig and J.S.Marron}\\
	Department of Statistics and Operations Research\\
	The University of North Carolina at Chapel Hill\\}
\date{\vspace{-5ex}}
\begin{document}
\maketitle

\begin{abstract}
		Integrative analysis of disparate data blocks measured on a common set of experimental subjects is one major challenge in modern data analysis. This data structure naturally motivates the simultaneous exploration of the joint and individual variation within each data block resulting in new insights. For instance, there is a strong desire to integrate the multiple genomic data sets in The Cancer Genome Atlas (TCGA) to characterize the common and also the unique aspects of cancer genetics and cell biology for each source. In this paper we introduce Non-iterative Joint and Individual Variation Explained (Non-iterative JIVE), capturing both joint and individual variation within each data block. This is a major improvement over earlier approaches to this challenge in terms of a new conceptual understanding, much better adaption to data heterogeneity and a fast linear algebra computation. Important mathematical contributions are the use of score subspaces as the principal descriptors of variation structure and the use of perturbation theory as the guide for variation segmentation. This leads to a method which is robust against the heterogeneity among data blocks without a need for normalization. An application to TCGA data reveals different behaviors of each type of signal in characterizing tumor subtypes. An application to a mortality data set reveals interesting historical lessons. 
\end{abstract}

\textbf{keywords}: Data Integration, \ Variation decomposition, \ Singular Value Decomposition, \ Principal Angel, \ Perturbation theory, \ Heterogeneity.

	\section{Introduction}
	\label{c:jive-s:intro}
	A major challenge in modern data analysis is data integration, combining diverse information from disparate data sets measured on a common set of experimental subjects. A unified and insightful understanding of the set of data blocks is expected from simultaneously exploring the joint variation representing the inter-block associations and the individual variation specific to each block. 
	
	\citet{lock2013joint} formulated this challenge into a matrix decomposition problem. Each data block is decomposed into three matrices modeling different types of variation, including a low-rank approximation of the joint variation across the blocks, low-rank approximations of the individual variation for each data block, and residual noise. Definitions and constraints were proposed for the joint and individual variation together with a method named \emph{JIVE} for obtaining a target decomposition. 
	
	The method JIVE developed a promising framework for studying multiple data matrices. However, the concepts of joint and individual variation were neither fully understood nor well defined. That lack of understanding of variation led to problems in computation. The~\citet{lock2013joint} algorithm was iterative (thus slow) and had no guarantee of achieving a solution that satisfied the definitions of JIVE. Another drawback of that approach includes a need for arbitrary normalization of the data sets which can be hard to choose in some complicated contexts. The example in Figure~\ref{fig:jive:toyoldjiveoutput} below shows these can be serious issues. A related algorithm was developed by~\citet{zhou2015group}, which consider a JIVE type decomposition as a quadratic optimization problem with restrictions to ensure identifiability. But it still requires iterations which might take longer computation time than the~\citet{lock2013joint} algorithm. Besides, neither~\citet{zhou2015group} nor~\citet{lock2013joint} provides statistical guarantee for the selection of joint components. 

	A novel solution is proposed here for addressing this matrix decomposition problem. It provides a relatively very efficient non-iterative algorithm ensuring an identifiable decomposition and also an insightful new interpretation of extracted variation structure. The key insight is the use of row spaces, i.e. a focus on scores, as the principal descriptor of the joint and individual variation, assuming columns are the $n$ data objects, e.g. vectors of measurements on patients. This focuses the methodology on variation patterns across data objects, e.g. patient signatures, which gives straightforward definitions of the components and thus provides identifiability. These variation patterns are captured by the \emph{row patterns} living in the row space, defined as \emph{score subspaces} of $\mathbb{R}^n$. Segmentation of joint and individual variation is based on studying the relationship between these score subspaces and using perturbation theory to quantify noise effects~\citep{stewart1990matrix}. 
	
	Using score subspaces to describe variation contained in a matrix not only empowers the interpretation of analysis but also improves the correctness and efficiency of the algorithm. An identifiable decomposition can now be obtained with all definitions and constraints satisfied. Moreover, the selection of one tuning parameter for~\citet{zhou2015group} to distinguish joint and individual variation is eliminated based on theoretical justification using perturbation theory. A consequence is a fast linear algebra based algorithm which no longer requires any iteration. The algorithm achieves an overall speedup factor around $16$ compared with JIVE, when analyzing the real data set introduced in Section~\ref{c:jive-s:intro-subsec:data}. A further benefit of this new approach is that a very problematic data normalization to handle data scaling and widely differing numbers of features is no longer needed as variation patterns are now quantified by score subspaces.
	
	Other methods that aim to study joint variation patterns and/or individual variation patterns have also been developed. \citet{westerhuis1998analysis} discusses two types of methods. One main type extends the traditional Principal Component Analysis (PCA), such as Consensus PCA and Hierarchical PCA first introduced by \citet{wold1987multi, wold1996hierarchical}. An overview of extended PCA methods is discussed in~\citet{smilde2003framework}. This type of method computes the block scores, block loadings, global loadings and global scores based on an iterative procedure. The other main type of method are extensions of Partial Least Squares (PLS)~\citep{wold1985partial} or Canonical Correlation Analysis (CCA)~\citep{hotelling1936relations} that seek associated patterns between the two data blocks by maximizing covariance/correlation. For example,~\citet{wold1996hierarchical} introduced multi-block PLS and hierarchical PLS (HPLS) and~\citet{trygg2003o2} proposed \emph{O2-PLS} to better reconstruct joint signals by removing structured individual variation.
	
	A connection between extended PCA and extended PLS methods is discussed in~\citet{hanafi2011connections}. Both types of methods provide an integrative analysis by taking the inter-block associations into account. These papers make the recommendations to use normalization to address potential scale heterogeneity, including normalizing by the Frobenius norm, or the largest singular value of each data block etc. However, there are no consistent criteria for normalization and some of these methods have convergence problems. An important point is that none of these approaches provide simultaneous decomposition highlighting joint and individual modes of variation with the goal of contrasting these to reveal new insights.
	
	\subsection{Practical Motivation}
	\label{c:jive-s:intro-subsec:data}
	Simultaneous variation decomposition has been useful in many practical applications, e.g. cancer genomic research. For example, \citet{lock2013bayesian}, \citet{kuhnle2011integration}, \citet{mo2013pattern} performed integrative clustering on multiple sources to reveal novel and consistent subtypes based on understanding of joint and individual variation. Other types of application include analysis of multi-source metabolomic data~\citep{kuligowski2015analysis}, extraction of commuting patterns in railway networks~\citep{jere2014extracting}, recognition of brain–computer interface~\citep{zhang2015ssvep} etc.
	
	The Cancer Genome Atlas (TCGA)~\citep{cancer2012comprehensive} provides a prototypical example for the application of JIVE. TCGA contains disparate data types generated from high-throughput technologies. Integration of these is fundamental for studying cancer on a molecular level. As a concrete example, we analyze gene expression, copy number variations, reverse phase protein arrays (RPPA) and gene mutation for a set of $616$ breast cancer tumor samples. For each tumor sample, there are measurements of $16615$ gene expression features, $24174$ copy number variations features, $187$ RPPA features and $18256$ mutation features. These data sources have very different dimensions and scalings.
	
	The tumor samples are classified into four molecular subtypes: Basal-like, HER2, Luminal A and Luminal B. An integrative analysis targets the association among the features of these four disparate data sources that jointly quantify the differences between tumor subtypes. In addition, identification of driving features for each source and subtype is obtained from studying loadings. 
	
	\subsection{Toy Example}
	\label{c:jive-s:intro-subsec:toy}
\begin{figure}[htp!]
		\begin{center}
			\vspace{0.1in}	
			\includegraphics[scale = 0.55]{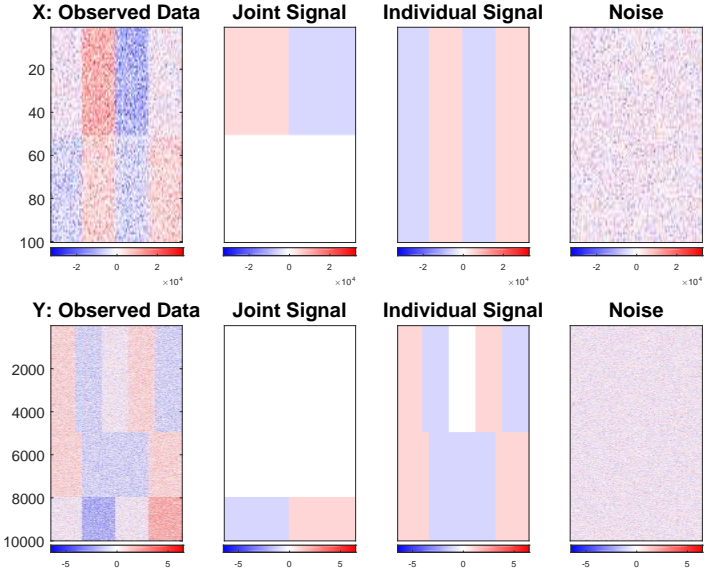}
			\caption[Data blocks $X$ and $Y$ in the toy example]{Data blocks $X$ (top) and $Y$ (bottom) in the toy example. The left panels present the observed data matrices which are a sum of the signal and noise matrices depicted in the remaining panels. Scale is indicated by color bars at the bottom of each sub-plot. These structures are challenging to capture using conventional methods due to very different orders of magnitude and numbers of features.}
			\label{fig:jive:toyrawdata}
		\end{center}
	\end{figure}

	A toy example provides a clear view of multiple challenges brought by potentially very disparate data blocks. This toy example has two data blocks, $X$ ($100\times 100$) and $Y$ ($10000\times 100$), with patterns corresponding to joint and individual structures. Figure~\ref{fig:jive:toyrawdata} shows colormap views of these matrices, with the value of each matrix entry colored according to the color bar at the bottom of each subplot. The signals have row mean $0$. Therefore mean centering is not necessary in this case. A careful look at the color bar scalings shows the values are almost 4 orders of magnitude larger for the top matrices. Each column of these matrices is regarded as a common data object and each row is considered as one feature. The number of features is also very different as labeled in the y-axis. Each of the two raw data matrices, $X$ and $Y$ in the left panel of Figure~\ref{fig:jive:toyrawdata}, is the sum of joint, individual and noise components shown in the other panels. 
	
	The joint variation for both blocks, second column of panels, presents a contrast between the left and right halves of the data matrix, thus having the same rank one score subspace. If for example the left half columns were male and right half were female, this joint variation component can be interpreted as a contrast of gender groups which exists in both data blocks for those features where color appears. 
	
	The $X$ individual variation, third column of panels, partitions the columns into two groups of size $50$ that are arranged so the row space signature is orthogonal to that of the joint score subspace. The individual signal for $Y$ contains two variation components, each driven by the half of the features. The first component, displayed in the first $5000$ rows, partitions the columns into three groups. The other component is driven by the bottom half of the features and partitions the columns into two groups, both with row spaces orthogonal to the joint. Note that these two individual score subspaces for $X$ and $Y$ are different but not orthogonal. The smallest angle between the individual subspaces is $48$\textdegree. 
	
	This example presents several challenging aspects, which also appear in real data sets such as TCGA. One is that the values of the features are orders of magnitude different between $X$ and $Y$. There are two standard approaches to handle this, both having drawbacks. Feature by feature normalization loses information in $X$ because $Y$ has so many more features. Total power normalization tends to underweight the signal in $Y$ because each feature then receives too little weight.    
	
	The noise matrices, the right panel of Figure~\ref{fig:jive:toyrawdata}, are standard Gaussian random matrices (scaled by 5000 for X) which generates a very noisy context for both data blocks and thus a challenge for analysis, as shown in the left panels of Figure~\ref{fig:jive:toyrawdata}.

\begin{figure}[htp!]
		\centering
		\vspace{0.1in}
		\includegraphics[scale = 0.55]{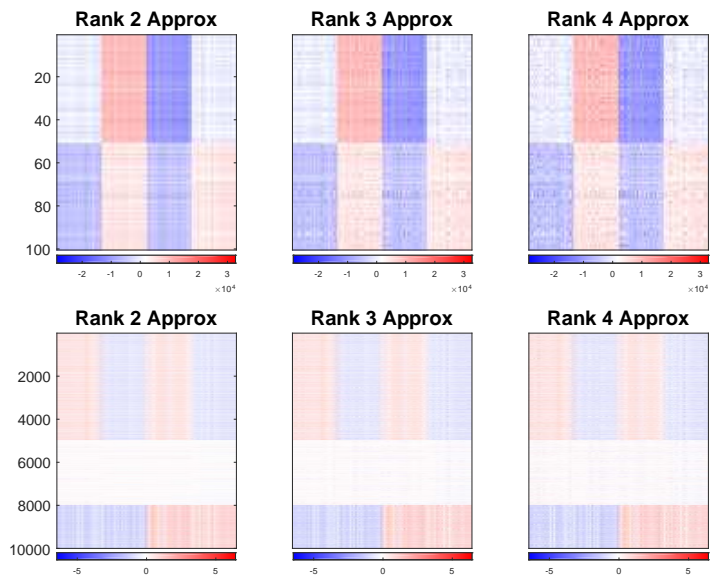}
		\caption[SVD approximations of concatenated toy data blocks]{Shows the concatenation SVD approximation of each block for rank 2 (left), 3 (center) and 4 (right). Although block $X$ has a relatively accurate approximation when the rank is chosen as 2, the individual pattern in block $Y$ has never been captured due to the heterogeneity between $X$ and $Y$.}
		\label{fig:jive:toysvdoutput}
	\end{figure}

	A first attempt at integrative analysis can be done by concatenating $X$ and $Y$ on columns and performing a singular value decomposition on this concatenated matrix. Figure~\ref{fig:jive:toysvdoutput} shows the results for $3$ choices of rank. The rank $2$ approximation essentially captures the joint variation component and the individual variation component of $X$, but the $Y$ components are hard to interpret. The bottom $2000$ rows show the joint variation but the top half of $Y$ reveals signal from the individual component of $X$.  One might hope that the $Y$ individual components would show up in the rank $3$ and rank $4$ approximations. However, because the noise in the $X$ matrix is so large, a random noise component from $X$ dominates the $Y$ signal, so the important latter component disappears from this low rank representation. In this example, this naive approach completely fails to give a meaningful joint analysis. 

\begin{figure}[htp!]
		\centering
		\includegraphics[scale = 0.55]{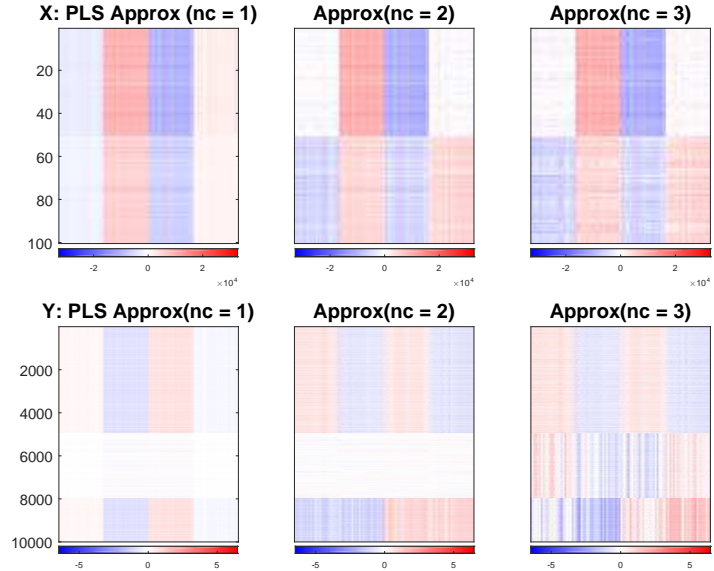}
		\caption[PLS approximations of the toy data]{PLS approximations of each block for numbers of components as 1 (left), 2 (center) and 3 (right). PLS fails to distinguish the joint and individual variation structure.}
		\label{fig:jive:toyplsoutput}
	\end{figure}

	PLS and CCA might be used to address the magnitude difference in this example and capture the signal components. However, they target common relationships between two data matrices and therefore are not able to simultaneously extract and distinguish the two types of variation. Figure~\ref{fig:jive:toyplsoutput} presents the PLS approximations with different numbers of components selected. PLS completely fails to separate joint and individual components. Instead it provides mixtures of the joint, and some of the individual components. Increasing the rank of the PLS approximation only includes more noise. 
	
\begin{figure}[htp!]
		\centering
		\vspace{0.1in}
		\includegraphics[scale = 0.55]{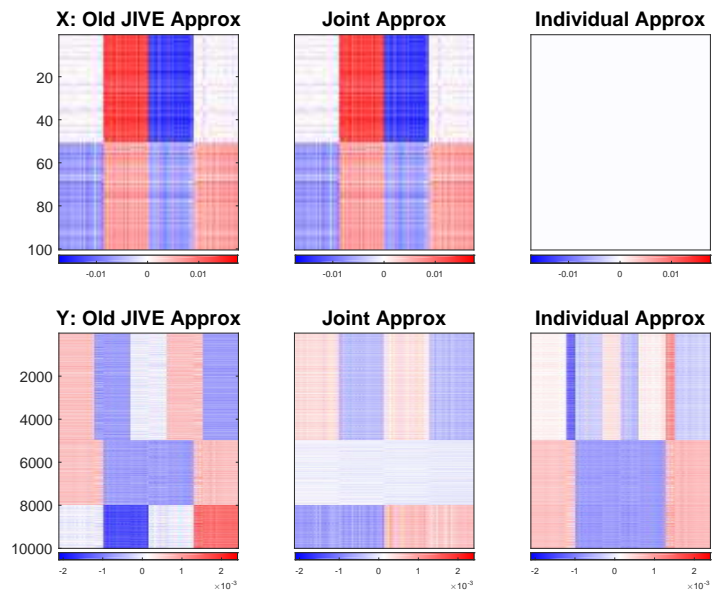}
		\caption[The old JIVE approximation of the toy data]{ The \citet{lock2013joint} JIVE method approximation of the data blocks $X$ and $Y$ in the toy example are shown in the first panel of figures. The joint matrix approximations (middle panel) incorrectly contain the individual component of $X$ caused by the problematic algorithm and inappropriate normalization.}
		\label{fig:jive:toyoldjiveoutput}
	\end{figure}

	The \citet{lock2013joint} method, called old JIVE here, is applied to this toy data set. The left panel of Figure~\ref{fig:jive:toyoldjiveoutput} shows a reasonable JIVE approximation of the total signal variation within each data block. However, the \citet{lock2013joint} method gives rank $2$ approximations to the joint matrices shown in the middle panel. The approximation consists of the real joint component together with the individual component of $X$. Following this, the approximation of the $X$ individual matrix is a zero matrix and a wrong approximation of the $Y$ individual matrix is obtained shown in the top half of the right panel. We speculate that failure to correctly apportion the joint and individual variation is caused by either the iterative algorithm that does not guarantee the JIVE conditions, and/or the Frobenius normalization of the individual components. 
	
\begin{figure}[htp!]
		\centering
		\vspace{0.1in}
		\includegraphics[scale = 0.55]{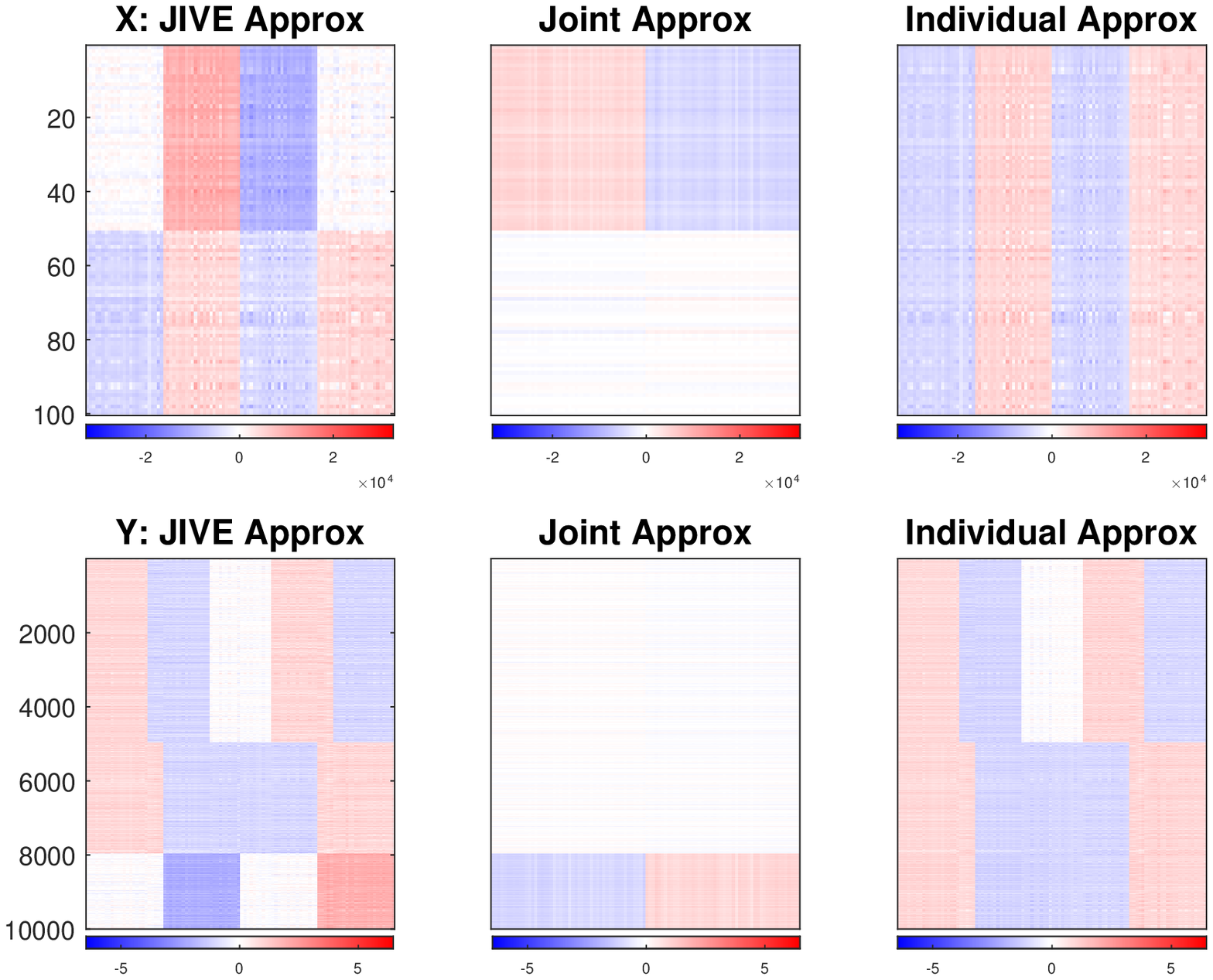}
		\caption[Non-iterative JIVE approximation of the toy data]{ Our JIVE method approximation of the data blocks $X$ and $Y$ in the toy example are shown in the first column of figures, with the joint and individual signal matrices depicted in the remaining columns. Both quite diverse types of variations are well captured for each data block by new proposed JIVE.}
		\label{fig:jive:toyjiveoutput}
	\end{figure}

	The left panel of Figure~\ref{fig:jive:toyjiveoutput} shows our JIVE approximation of each data block which well captures the signal variations within both $X$ and $Y$. What's more, our method correctly distinguishes the types of variation showing its robustness against heterogeneity across data blocks. The approximations of both joint and individual signal are depicted in the remaining panels. 
	
	The rest of this paper is organized as follows. Section~\ref{c:jive-s:method} describes the population model and the estimation approach. Results of application to a TCGA breast cancer data set and a mortality data set are presented in Section~\ref{c:jive-s:datanalaysis}. 

	\section{Proposed Method}
	\label{c:jive-s:method}
	In this section the details of the new proposed JIVE are discussed. The population model is proposed in Sections~\ref{c:jive-s:method-subsec:model} and~\ref{c:jive-s:method-subsec:noise}. The theoretical foundations based on matrix perturbation theory from linear algebra~\citep{stewart1990matrix} are given in Section~\ref{c:jive-s:method-subsec:theoretical}. These theoretical results motivate our estimation approach which is proposed in Section~\ref{c:jive-s:method-subsec:estimate}.
	
	\subsection{Population Model - Signal}
	\label{c:jive-s:method-subsec:model}
	Matrices $\{X_k, \ k = 1, \ldots, K\}$ ($d_k \times n$) are a set of data blocks for study. The columns are regarded as data objects, one vector of measurements for each experimental subject, while rows are considered as features. All $X_k$ therefore have the same number of columns and perhaps a different number of rows.
	
	Each $X_k$ is modeled as low rank signals $A_k$ perturbed by additive noise matrices $E_k$. Each low rank signal $A_k$ is the sum of two matrices containing joint and individual variation, denoted as $J_k$ and $I_k$ respectively for each block 
	\begin{equation}
	\begin{bmatrix}
	X_1 \\
	X_2 \\ 
	\vdots \\
	X_K \\
	\end{bmatrix} = \begin{bmatrix}
	A_1 \\
	A_2 \\ 
	\vdots \\
	A_K \\
	\end{bmatrix}  + \begin{bmatrix}
	E_1 \\
	E_2 \\ 
	\vdots \\
	E_K \\
	\end{bmatrix}  = \begin{bmatrix}
	J_1 \\
	J_2 \\ 
	\vdots \\
	J_K \\
	\end{bmatrix}  + \begin{bmatrix}
	I_1 \\
	I_2 \\ 
	\vdots \\
	I_K \\ 
	\end{bmatrix}  + \begin{bmatrix}
	E_1 \\
	E_2 \\ 
	\vdots \\
	E_K \\
	\end{bmatrix}.
	\end{equation}
	Our approach focuses on \emph{score vectors}, e.g. patient \textit{signatures}, which are determined by the \textit{row patterns} living in the row space, $\mathbb{R}^n$. These row patterns are essentially represented by the right basis vectors of appropriate SVDs. These score vectors generate the \emph{score subspace}($\subset \mathbb{R}^n$). Therefore, the matrices capturing joint variation i.e. joint matrices are defined as sharing a common score subspace denoted as $\row(J)$
	$$\row(J_k) = \row(J), \quad k = 1, \ldots, K.$$
	The individual matrices are individual in the sense that the intersection of their score subspaces is the zero vector space, i.e.
	$$\bigcap\limits_{k=1}^{K} \row(I_k) = \{\vec{0}\}, \quad k = 1, \ldots, K.$$
	This means that there is no non-trivial common row pattern living in every individual score subspaces across blocks. To ensure an identifiable variation decomposition, orthogonality between the score subspaces of matrices containing joint and individual variation is assumed. In particular, $\row(J) \perp \row(I_{k})$, $k = 1, \ldots, K$. Note that orthogonality between individual matrices $\{I_k, \ k = 1, \ldots, K\}$ is \textit{not} assumed as it is not required for the model to be uniquely determined. The relationship between individual matrices, to some extent, has an impact on the estimation accuracy which will be discussed in Section~\ref{c:jive-s:method-subsec:estimate}.
	
	Under these assumptions, the model is identifiable in the sense:
	\begin{theorem}
		\label{lemma-exist}
		Given a set of matrices $\{A_k, \ k = 1, \ldots, K\}$, there are unique sets of matrices $\{J_k, \ k = 1, \ldots, K\}$, and $\{I_k, \ k = 1, \ldots, K\}$ so that:
		\begin{enumerate}
			\item $A_k$ = $J_k + I_k$, $k = 1, \ldots, K$
			\item $\row(J_k) = \row(J)$, $k = 1, \ldots, K$
			\item $\row(J) \perp \row(I_k)$, $k = 1, \ldots, K$
			\item $\bigcap\limits_{k=1}^{K} \row(I_k) = \{\vec{0}\}.$
		\end{enumerate}
	\end{theorem}
	The proof is provided in the Appendix. This model has enhanced the matrix decomposition idea proposed in \citet{lock2013joint} by providing a clearer mathematical framework and precise understanding of the different types of variation. In particular, \citet{lock2013joint} imposed rank constraints on the joint matrices i.e. $rank(J_k)$ are the same for all data blocks but did not clearly formulate the definition of a common row pattern. Furthermore, the orthogonality constraint was formulated on matrices instead of score subspaces i.e. $J_kI_{k}^{T} = \mathbf{0}$, which tended to obscure the role of row spaces in defining variation structure. An unnecessary orthogonality among individual matrices was further suggested, although not explicitly enforced in the estimation, for ensuring a well defined decomposition. 
	
	\subsection{Population Model - Noise}
	\label{c:jive-s:method-subsec:noise}
	The additive noise matrices are assumed to follow an isotropic error model where the energy of projection is invariant to direction in both row and column spaces. Important examples include the multivariate standard normal distribution and the multivariate student t--distribution~\citep{kotz2004multivariate}. The singular values of each noise matrix are assumed to be smaller than the smallest singular values of each signal to give identifiability. 
	
	The assumption on the noise distribution here is less strong than the classical i.i.d. Gaussian random matrix, and only comes into play when determining the number of joint components. Other than that, the estimation approach given in Section~\ref{c:jive-s:method-subsec:theoretical} reconstructs each signal matrix based on SVD and thus is quite robust against the error distribution.
	
	\subsection{Theoretical Foundations}
	\label{c:jive-s:method-subsec:theoretical}
	The main challenge is segmentation of the joint and individual variation in the presence of noise which individually perturbs each signal. Let $\{\tilde{A}_k, k = 1, \ldots, K\}$ be noisy approximations of $\{A_k, k = 1, \ldots, K\}$ respectively. The subspaces of joint variation within the approximations $\tilde{A}_k$, while expected to be similar, are no longer exactly the same due to noise. If some subspaces of $\{\tilde{A}_k, k = 1, \ldots, K\}$ are very close, they can be considered as estimates of the common score subspace under different perturbations. Application of the results of the \emph{Generalized $\sin\theta$ Theorem}~\citep{wedin1972perturbation} is proposed to decide when a set of subspaces are close enough to be regarded as estimates of the joint score space. Based on this theorem, the number of joint components can be determined resulting in an appropriate segmentation.
	
	Take the approximation $\tilde{A}_k$ of $A_k$ as an example of perturbation of each matrix's score space. For consistency with the Generalized $\sin\theta$ Theorem, a notion of distance between theoretical and perturbed subspaces is defined as a measure of perturbation. Let $\mathcal{Q}$, $\mathcal{\tilde{Q}}$ be the $l$ dimensional score subspaces of $\mathbb{R}^{n}$ respectively for the matrix $A_k$ and its approximation $\tilde{A}_k$. The corresponding symmetric projection matrices are $P_{\mathcal{Q}}$ and $P_{\mathcal{\tilde{Q}}}$. The distance between the two subspaces is defined as the difference of the projection matrices under the $L^2$ operator norm i.e $\rho(\mathcal{Q}, \mathcal{\tilde{Q}}) = \|P_{\mathcal{Q}} - P_{\mathcal{\tilde{Q}}}\|$~\citep{stewart1990matrix}.
	
	An insightful understanding of this defined distance $\rho(\mathcal{Q}, \mathcal{\tilde{Q}})$ comes from a principal angle analysis~\citep{jordan1875essai, hotelling1936relations} of the subspaces $\mathcal{Q}$ and  $\mathcal{\tilde{Q}}$. Denote the principal angles between $\mathcal{Q}$ and $\mathcal{\tilde{Q}}$ as $\Theta(\mathcal{Q}, \mathcal{\tilde{Q}}) = \{\theta_1, \ldots, \theta_l\}$ with $\theta_1 \geq \theta_2 \ldots \geq \theta_l$. The distance $\rho$ is equal to the sine of the maximal principal angle, i.e. $\sin\theta_{1}$. This suggests that the largest principal angle between two subspaces can indicate their closeness, i.e. distance. Under a slight perturbation, the largest principal angle between $\mathcal{Q}$, a theoretical subspace, and $\mathcal{\tilde{Q}}$, its perturbed subspace, is expected to be small.
	
	The distance $\rho(\mathcal{Q}, \mathcal{\tilde{Q}})$ can be also written as 
	$$
	\rho(\mathcal{Q}, \mathcal{\tilde{Q}}) = \|(I-P_{\mathcal{Q}})P_{\mathcal{\tilde{Q}}}\| = \|(I-P_{\mathcal{\tilde{Q}}})P_{\mathcal{Q}}\|
	$$
	which brings another useful understanding of this definition. It measures the relative deviation of the signal variation from the theoretical subspace. Accordingly, the similarity/closeness between the subspaces and its perturbation can be written as $\|P_{\mathcal{Q}}P_{\mathcal{\tilde{Q}}}\|$ and is equal to the cosine of the maximal principal angle defined above, i.e. $\cos\theta_{1}$. Hence, $\sin^2\theta_{1}$ indicates the percentage of signal deviation and $\cos^2\theta_{1}$ tells the percentage of remaining signal in the theoretical subspace. 
	
	% Wedin theorem
	The generalized $\sin\theta$ theorem provides a bound for the distance between a subspace and its perturbation, e.g., the subspaces $\mathcal{Q}$ and $\mathcal{\tilde{Q}}$. This bound quantifies how the theoretical subspace $\mathcal{Q}$ is affected by noise. In particular,
	
	\begin{theorem}[The Generalized $\sin\theta$ Theorem (Wedin, 1972)]
		\label{thm-wedin}
		Signal matrix $A_k$ is perturbed by additive noise $E_k$. Let $\theta_{k}$ be the largest principal angle for the subspace of signal $A_k$ and its approximation $\tilde{A}_k$. Denote the SVD of $\tilde{A}_k$ as $\tilde{U}_k\tilde{\Sigma}_k\tilde{V}_k^{T}$. The distance between the subspaces of $A_k$ and $\tilde{A}_k$, $\rho(\mathcal{Q}, \mathcal{\tilde{Q}})$ i.e. sines of $\theta_{k}$, is bounded
		\begin{equation}
		\rho(\mathcal{Q}, \mathcal{\tilde{Q}}) = \sin\theta_{k} \leq \frac{\textnormal{max}(\|E_k\tilde{V}_k\|, \|E_k^{T}\tilde{U}_k\|)}{\sigma_{\textnormal{min}}(\tilde{\Sigma}_k)},
		\end{equation}
		where $\sigma_{\textnormal{min}}(\tilde{\Sigma}_k)$ is the smallest singular value of $\tilde{A}_k$.
	\end{theorem}
	
	This bound measures how far the perturbed space can be away from the theoretical one. The deviation is bounded by the maximal value of noise energy on the column and row spaces and also the smallest signal singular values. This is consistent with the intuition that a deviation distance, i.e. a largest principal angle, is small when the signal is strong and perturbations are weak.
	
	Notice that the bound in Theorem~\ref{thm-wedin} is applicable but cannot be directly used for data analysis since the error matrices $E_k$ are not observable. As the error matrices are assumed to be isotropic, we propose to re-sample noisy directions from the residuals of the low rank approximations. The $L^2$ norm of these error related terms can thus be estimated by projecting the observed data onto the subspace spanned by re-sampled directions. This re-sampling based method can also provide confidence intervals for these perturbation bounds. More details of estimating the perturbation bound will be discussed in Section~\ref{c:jive-s:method-subsec:estimate}.
	
	\subsection{Estimation Approach}
	\label{c:jive-s:method-subsec:estimate}
	The algorithm uses SVD as a building block to find an estimate of the targeted decomposition. A three-step algorithm is outlined below.
	
	\begin{enumerate}
		\item Obtain an initial estimate of the signal score space of each data block by thresholding the singular values.
		\item Extract the joint score space from the signal score spaces using a threshold derived from Theorem~\ref{thm-wedin}.
		\item Decompose each data matrix into joint and individual variation matrices using projections onto the score space in Step 2.
	\end{enumerate}
	
	As a basic illustration for each step we use the toy example described in Section~\ref{c:jive-s:intro}. Details for each step appear in the following subsections.
	
\begin{figure}[htb!]
		\vspace{0.1in}
		\begin{minipage}[b]{0.5\textwidth}
			\centering
			\includegraphics[scale=0.4]{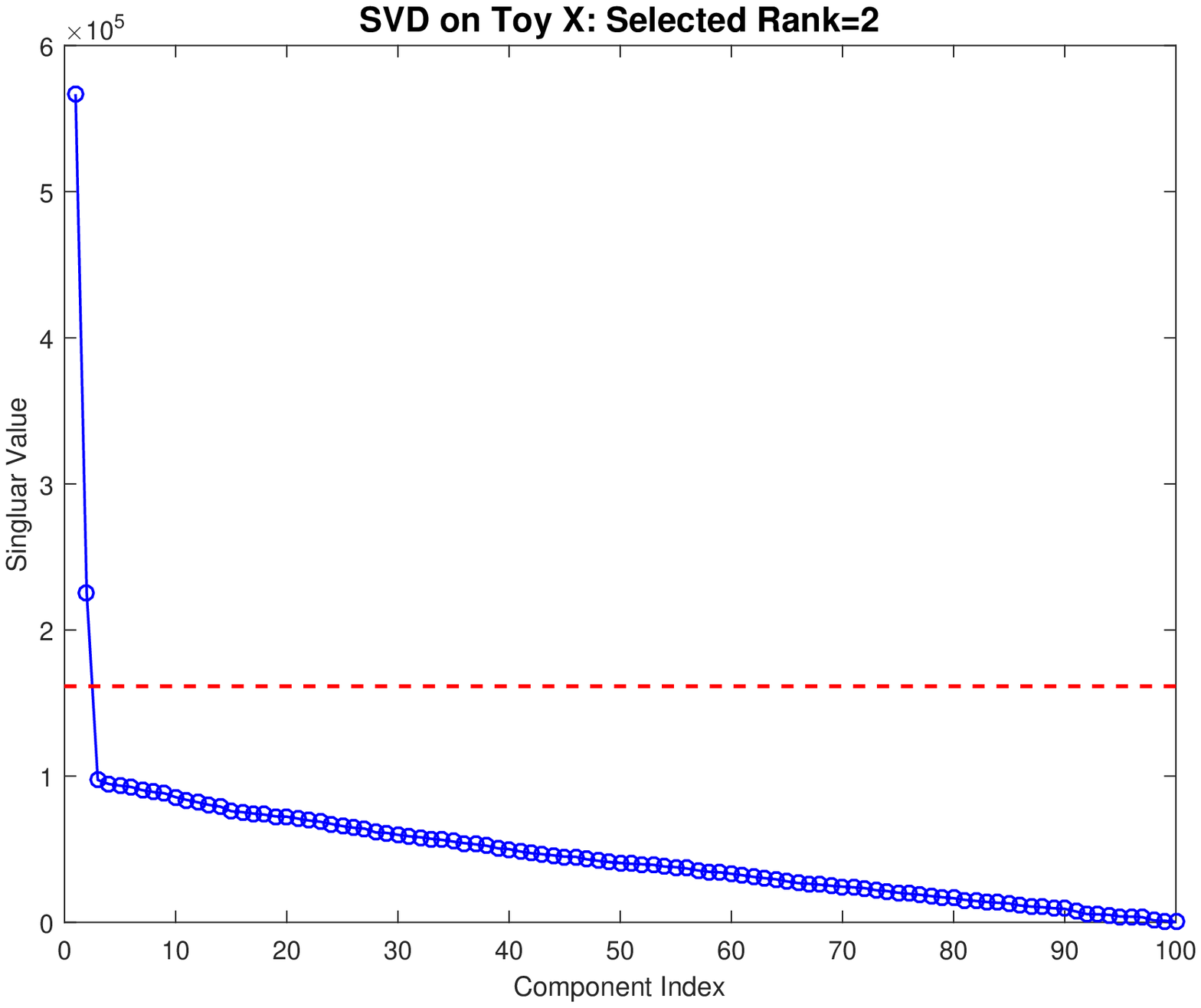}
			\vspace{4ex}
		\end{minipage}%%
		\begin{minipage}[b]{0.5\textwidth}
			\centering
			\includegraphics[scale=0.4]{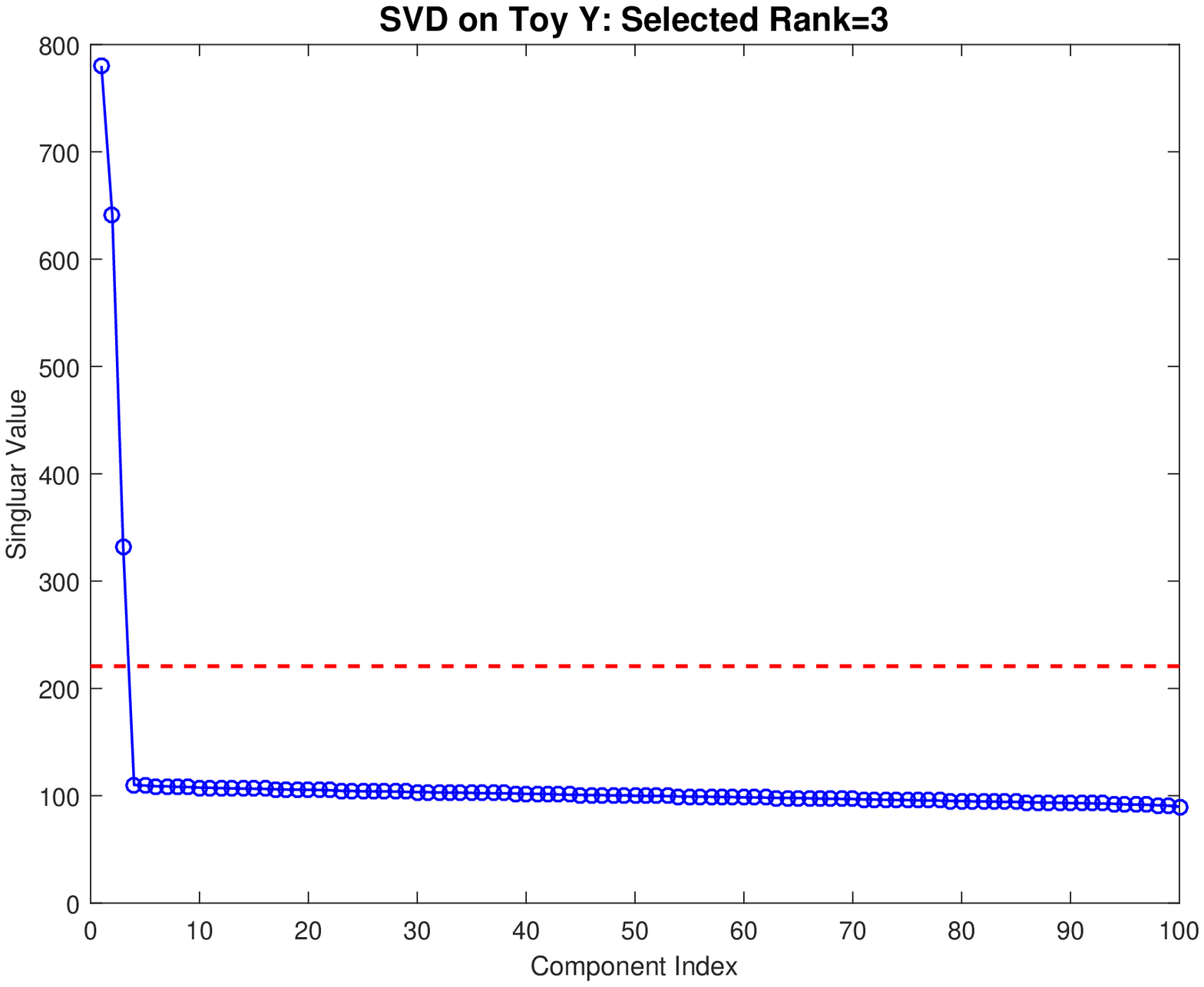}
			\vspace{4ex}
		\end{minipage}
		\caption[Scree plots for the toy data]{Scree plots for the toy data sets $X$ (left) and $Y$ (right). Both plots display the singular values associated with a component in descending order versus the index of the component. The components with singular values above the dashed red threshold line are regarded as the initial signal components in the first step of JIVE.}
		\label{fig:toy1singularvalue}
	\end{figure}

	\textbf{Signal Space Initial Extraction: }
	\label{subsubsec:step1}
	Even though the signal components $\{A_k, \ k = 1, \ldots, K\}$ are low rank, the data matrices $\{X_k, \ k = 1, \ldots, K\}$ are usually of full rank due to corruption by noise. SVD works as a signal extraction device in this step, keeping components with singular values greater than selected thresholds individually for each data block. These thresholds are selected using a multi-scale perspective. For example, by finding relatively big jumps in a scree plot. Figure~\ref{fig:toy1singularvalue} shows the scree plots of each data block for the toy example in Section~\ref{c:jive-s:intro}. The left scree plot for $X$ suggests a selection of rank as $2$ and the right one for $Y$ suggests the rank being $3$, since in both cases those components stand out while the rest of the singular values decay slowly showing no clear jump.
	
	Let $\{\tilde{r}_k, \ k = 1, \ldots, K\}$ be the initial estimates of the signal ranks $\{r_k, \ k = 1, \ldots, K\}$. In the toy example ${\tilde{r}_1} = 2$ (for $X$) and ${\tilde{r}_2} = 3$ (for $Y$). Each data block has a low rank approximation, $\tilde{A}_k$, which is the initial estimate of the signal matrix $A_k$, $k = 1, \ldots, K$. The estimate is decomposed as
	\begin{equation}
	\label{equ:step1}
	\tilde{A}_k = \tilde{U}_k\tilde{\Sigma}_k\tilde{V}_k^{T}
	\end{equation}
	where $\tilde{U}_{k}$ contains the left singular vectors corresponding to the largest $\tilde{r}_k$ singular values respectively for each data block. The initial estimate of the signal score space, denoted as $\row(\tilde{A}_{k})$, is spanned by the right singular vectors in $\tilde{V}_{k}$.
	
	\textbf{Score Space Segmentation of Two-Block: }
	\label{subsubsec:step2multi}
	For a clear introduction of the basic idea of score space segmentation, the two-block special case ($K=2$) is first studied. The goal is to use the low rank approximations $\tilde{A}_k$ from equation~(\ref{equ:step1}) to obtain estimates of the common joint and individual score subspaces. Due to the presence of noise, the components of $\row(\tilde{A}_{k})$ corresponding to the underlying joint space, no longer are the same, but should have a relatively small angle. Similarly, the components corresponding to the underlying individual spaces are expected to have a relatively large angle. This motivates the use of principal angle analysis to separate the joint from the individual components. The following Lemma~\ref{lemma-bound} provides a bound on the largest allowable principal angle of the joint part of the initial estimated spaces. 
	
	\begin{lemma}
		\label{lemma-bound}
		Let $\phi$ be the largest principal angle between two subspaces that are each a perturbation of the common row space within $\row(\tilde{A}_1)$ and $\row(\tilde{A}_2)$. That angle is bounded by 
		\begin{equation}
		\sin\phi \leq \sin(\theta_{1} + \theta_{2})
		\end{equation}
		in which $\theta_{1}$ and $\theta_{2}$ are the angles given in Theorem~\ref{thm-wedin}.
	\end{lemma}
The proof is provided in the Appendix. As mentioned in Section~\ref{c:jive-s:method-subsec:theoretical}, the perturbation bounds of each $\theta_k$ require the estimation of terms $\|E_k\tilde{V}_k\|$, $\|E_k^{T}\tilde{U}_k\|$ for $k = 1, 2$. These terms are the measurements of energies of noise matrices projected onto the signal column and row spaces. Since an isotropic error model is assumed, the energy of the noise matrices in arbitrary directions are supposed to be equal. Denote $\tilde{V}_k^{\perp}$ ($n \times (\textnormal{min}(d_k,n)-\tilde{r}_k)$) and $\tilde{U}_k^{\perp}$ ($d_k \times (\textnormal{min}(d_k,n)-\tilde{r}_k)$) as the respective orthonormal bases of the row and column subspaces of the residual matrices from the low rank approximations in equation~(\ref{equ:step1}). Thus, we propose to resample noisy directions, i.e. column vectors, from the matrices $\tilde{V}_k^{\perp}$ and $\tilde{U}_k^{\perp}$. 

Take the term $\|E_k\tilde{V}_k\|$ as an example for illustration. Given the $\tilde{r}_k$ number of column vectors resampled from $\tilde{V}_k^{\perp}$, denoted as $V^{\star}$, the observed data block $X_k$ is projected onto the subspace spanned by $V^{\star}$, written as $X_kV^{\star}$. The distribution of the $L^2$ norm $\|X_kV^{\star}\|$ approximates the distribution of the unknown $\|E_k\tilde{V}_k\|$. Thus we resample 1000 of $\|X_kV^{\star}\|$ and use the quantiles to provide both a point estimate and a simulated confidence interval for $\|E_k\tilde{V}_k\|$. This can be similarly applied to $\|E_k^{T}\tilde{U}_k\|$ for $k = 1, 2$, resulting in a confidence interval for the perturbation bound. Typically the median is chosen as the estimate of the angle bound for exploratory analysis. This will result in at least $50\%$ confidence that all joint components are included. For certain cases that where minimal of joint components is desired, the $95^{th}$ percentile of these estimated terms can be used to derive a conservative angle threshold, resulting in at least $95\%$ confidence of finding all joint components. 
	
	The principal angles between $\row(\tilde{A}_{1})$ and $\row(\tilde{A}_{2})$  are computed by performing SVD on a concatenation of their right singular vector matrices~\citep{miao1992principal}, i.e.
	\begin{equation}
	M \triangleq \begin{bmatrix}
	\tilde{V}_{1}^{T} \\
	\tilde{V}_{2}^{T} \\
	\end{bmatrix} = U_{M}\Sigma_{M}V_{M}^{T},
	\end{equation}
	where the singular values $\Sigma_M$ determine the principal angles, $\Phi(\row(\tilde{A}_1), \row(\tilde{A}_2))$ = $\{\phi_1, \ldots, \phi_l\}$ as
	\begin{equation}
	\label{equ: principalangle}
	\phi_i = \arccos((\sigma_{M,i})^2-1), \ i=1, \ldots, \min(\tilde{r}_1, \tilde{r}_2).
	\end{equation}
	
	Given a left singular vector $U_{M,i}$ denoted as $\vec{u}$, a pair of principal vectors $\{\vec{p}_i, \vec{q}_i\}$ in each subspace can be constructed by projecting $\tilde{V}_1$ and $\tilde{V}_2$ onto the vector $\vec{u}$. Denote $\vec{u}$ as the concatenation of $[\vec{u}_{1}; \vec{u}_{2}]$. Note that the length of $\vec{u}_{1}$ is equal to the number of columns of $\tilde{V}_1$ and similarly for the other part. The principal vectors in each subspace can be written as $\vec{p}_i = \tilde{V}_1\vec{u}_{1}$ and $\vec{q}_i = \tilde{V}_2\vec{u}_{2}$ respectively. The angle between the pair of principal vectors $\theta_i$ is equal to the principal angle computed from the singular value corresponding to $\vec{u}$.
	
	As seen in~\citet{miao1992principal}, the vector $\vec{v}_i$, the corresponding right singular vector of $V_{M}$, points in the same direction as the sum of principal vector pairs of each subspace. When the principal angle $\phi_i$ is smaller than the perturbation bound $\phi$, this right singular vector can be taken as an estimate of the theoretical joint direction to assure the definition of joint variation. 

This SVD decomposition can be understood as a tool that sorts pairs of directions within the two subspaces in increasing order of the angle between each pair. When the corresponding principal angle is smaller than the perturbation bound $\phi$, the pair of principal vectors can be considered as noisy versions of the same joint direction. Assume there are $\tilde{r}_J$ principal angles smaller than the bound $\phi$. The first $\tilde{r}_J$ singular vectors $\vec{v}_i$ are used as the natural orthonormal basis of the estimated joint score subspace. 

\begin{figure}[htb!]
		\vspace{0.1in}
		\begin{minipage}[b]{0.5\textwidth}
			\centering
			\includegraphics[scale = 0.4]{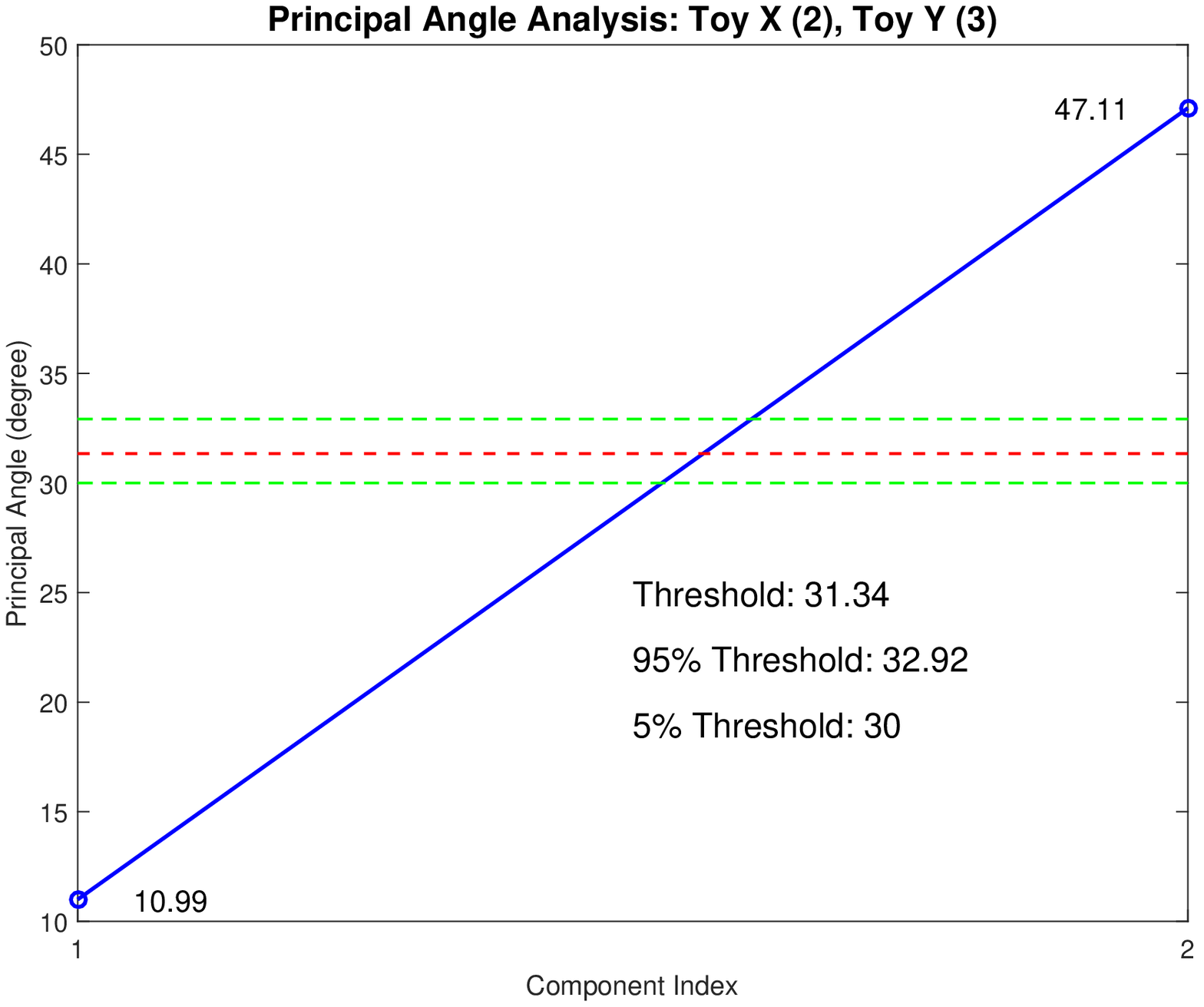}
		\end{minipage}%% 
		\begin{minipage}[b]{0.5\textwidth}
			\centering
			\includegraphics[scale = 0.4]{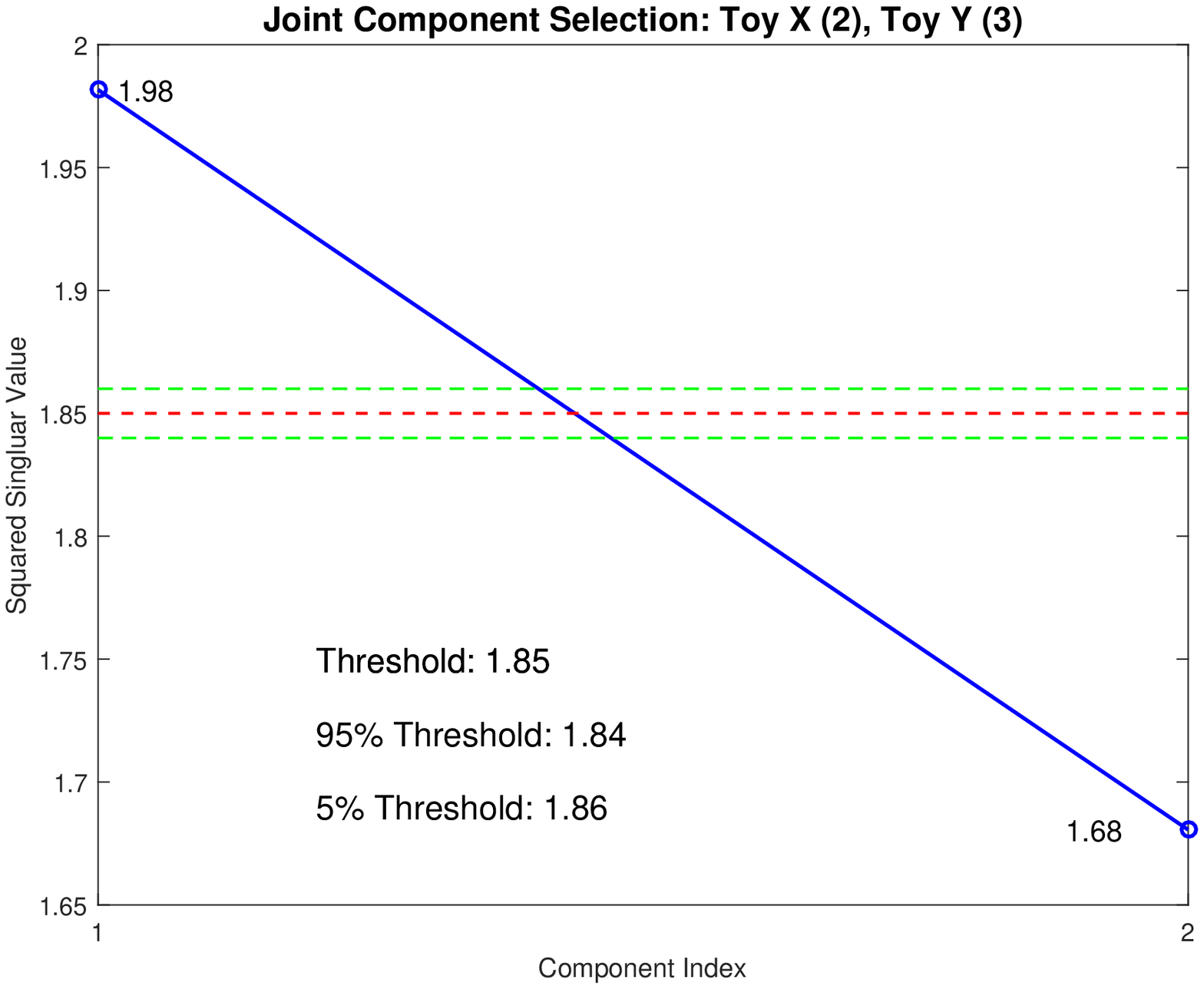}
		\end{minipage}%%
		\caption[Joint component selection]{Left panel: Principal angles between the initial estimates of signal row spaces. The bound for the largest angle is $31.29$ degree, suggesting the existence of one joint component. To indicate the uncertainty, the $5\%$ and $95\%$ one-sided confidence intervals of the angle threshold are also shown. Right panel: Squared singular values plot of the vertical concatenated matrix $M$ for the toy example. Both thresholds correctly capture the underlying structure of this toy example with the selection of one joint component.}
		\label{fig:toy1step2}
	\end{figure}

The left panel of Figure~\ref{fig:toy1step2} depicts the principal angles of the concatenated right singular vector matrices for the toy example in Section~\ref{c:jive-s:intro-subsec:toy}. Since the initial estimates of $r_x$ and $r_y$ are $2$ and $3$, there are only two potential components for joint variation. The associated principal angles between the initially estimated signal row spaces are labeled next to the first two components as $10.99$\textdegree~and $47.11$\textdegree.~The estimated bound on the principal angle in Lemma~\ref{lemma-bound} is $31.29$\textdegree~for this toy example. The $5\%$ and $95\%$ one-sided confidence intervals of the angle bound are $[-\infty, \ 30.00]$ and $[-\infty, \ 32.92]$ degrees. Each provides a respective $5\%$ and $95\%$ chance for including all the joint components. This provides a clear indication that the number of joint components should be $\tilde{r}_J=1$. The corresponding first right singular vector of $M$ will be taken as the joint score vector.

	\textbf{Score Space Segmentation of Multi-Block: }
	To generalize the above idea to more than two blocks, the key is to focus more on singular values than on angles in the equation~(\ref{equ: principalangle}). In other words, instead of finding an upper bound on an angle, we will focus on a lower bound on the remaining energy as expressed by the sum of the squared singular values. Hence, an analogous SVD will be used for studying the closeness of multiple initial signal score subspace estimates. 
	
	Similarly, for the vertical concatenation of right singular vector matrices $M \triangleq \{\tilde{V}_k^T, \ k = 1, \ldots, K\}$, SVD sorts the directions within these $K$ subspaces in increasing order of amount of deviation from the theoretical joint direction. The squared singular value $\sigma_{M,i}^2$ indicates the total amount of variation explained in the common direction $V_{M,i}^T$ in the score subspace of $\mathbb{R}^n$. A large value of $\sigma_{M,i}^2$ (close to $K$) suggests that there is a set of basis vectors within each subspace that are close to each other and thus are potential noisy versions of a common joint score vector. A threshold on singular values is needed to segment the joint components. This is done in Lemma~\ref{lemma-bound-multi}. 
	
	\begin{lemma}
		\label{lemma-bound-multi}
		Let $\theta_k$ be the bound on the principal angles between the theoretical subspace $\row(A_k)$ and its perturbation $\row(\tilde{A}_k)$ for $K$ data blocks from Theorem~\ref{thm-wedin}. The squared singular values ($\sigma_{M,i}^2$) corresponding to the estimates of joint components satisfy
		\begin{equation}
		\sigma_{M,i}^2 \geq K - \sum_{k=1}^{K}\sin^2\theta_{k} \geq K - \sum_{k=1}^{K}\big(\frac{\max(\|E_k\tilde{V}_k\|, \|E_k^{T}\tilde{U}_k\|)}{\sigma_{\min}(\tilde{\Sigma}_k)}\big)^2.
		\end{equation}
	\end{lemma}
	The proof is provided in the Appendix. This lower bound is independent of the variation magnitudes. This property gives some robustness against heterogeneity across each block when extracting joint variation information.   
	
As above, the terms $\|E_k\tilde{V}_k\|$, $\|E_k^{T}\tilde{U}_k\|$ are resampled to derive a point estimate and confidence interval for the threshold. As for the two-block case, if there were $\tilde{r}_J$ singular values selected, the first $\tilde{r}_J$ right singular vectors are used as the basis of the estimate of $\row(J)$.  

The right panel of Figure~\ref{fig:toy1step2} depicts the first $2$ singular values of the vertical concatenated matrix $M$ for the toy example. This is an analysis of the same data, but performed on the scale of squared singular values instead of principal angles. The associated squared singular values are labeled next to the these two component as $1.98$ and $1.68$. The estimated threshold, using median, is $1.85$ for the toy example. This threshold together with its $5\%$ and $95\%$ one sided confidence intervals, $[1.86, \ +\infty]$ and $[1.84, \ +\infty]$ respectively, suggest that the number of joint components $\tilde{r}_J$ should be $1$. 
	
	\textbf{Final Decomposition: }
	\label{subsubsec: decomp}
Based on the estimate of the joint row space, matrices containing joint variation in each data block can be reconstructed by projecting $X_{k}$ onto this estimated space. Define the matrix $\tilde{V}_{J}$ as $[\vec{v}_{M,1}, \ldots, \vec{v}_{M, \tilde{r}_J}]$, where $\vec{v}_{M,i}$ is the $i^{th}$ column in the matrix $V_M$. To ensure that all components continue to satisfy the identifiability constraints from Section~\ref{c:jive-s:method-subsec:noise}, we check that, for all the blocks, each $\|X_k\vec{v}_{M,i}\|$ is also above the corresponding threshold used in Step 1. If the constraint is not satisfied for any block, that component is removed from $\tilde{V}_{J}$. The removal can happen when there is a common joint structure in all but a few blocks. 

Denote $\hat{V}_J$ as the matrix $\tilde{V}_{J}$ after the removal and $\hat{r}_J$ as the final joint rank. The projection matrix onto the final estimated joint row space $\row(\hat{J})$ is $P_{J}=\hat{V}_{J}$. The estimates of joint variation matrices in each block are
\begin{equation*}
\hat{J}_{k} = X_{k}P_{J}, \quad k = 1, \ldots, K.
\end{equation*}
The row space of joint structure is orthogonal to the row spaces of each individual structure. Therefore, the original data blocks are projected to the orthogonal space of $\row(\hat{J})$. The projection matrix onto the orthogonal space of $\row(\hat{J})$ is $P_{J}^{\perp} = I-P_{J}$ and the projections of each data block are denoted as $X_{k}^{\perp}$ respectively for each block i.e.
\begin{equation*}
X_{k}^{\perp}=X_{k}P_{J}^{\perp}
\end{equation*}
Finally we threshold this projection by performing SVD on $\{X_{k}^{\perp}, \ k= 1, \ldots, K\}$. The components with singular values larger than the first thresholds from Section~\ref{subsubsec:step1} are kept as the individual components, denoted as $\{\hat{I}_{k}^{\perp}, \ k= 1, \ldots, K\}$. The remaining components of each SVD are regarded as an estimate of the noise matrices.

By taking a union of the estimated row spaces of each type of variation, denoted by $\oplus$, the estimated signal row spaces are
\begin{equation*}
\row(\hat{A}_k) = \row(\hat{J}) \oplus \row(\hat{I}_k)
\end{equation*}
with rank $\hat{r}_k = \hat{r}_{J} + \hat{r}_{I_k}$ respectively for $k = 1, \ldots, K$.

Due to this adjustment of directions of the joint components, these final estimates of signal row spaces may be different from those obtained in the initial signal extraction step. Note that even the estimates of rank $\hat{r}_k$ might also differ from the initial estimates $\tilde{r}_k$.

\section{Post JIVE Data Representation}
Given the variation decompositions of each data block, several types of post JIVE representations are available for exploring the joint and individual variation patterns. The estimates of joint matrices within each data block can be represented by SVD
\begin{equation}
\hat{J}_k = \hat{U}^{k}_J \hat{\Sigma}^{k}_J \hat{V}^{k}_J, \quad k = 1, \ldots, K
\end{equation}
in which $\hat{V}^{k}_J$ are the $\hat{r}_J \times n$ joint score matrices. Note that the singular values $\hat{\Sigma}^{k}_J$ can be completely different across $k$, since they are driven by the score variation pattern and can reflect very different amounts of variation between the blocks. The loading matrices $\hat{U}^{k}_J$ ($d_k \times \hat{r}_J$) respectively specify distinct $\hat{r}_J$-dimension loading subspaces of $\mathbb{R}^{d_k}$ for each block $k$.

There are three important matrix representations of the information in the joint score space, with differing uses in post JIVE analyses. 
\begin{enumerate}
	\item \emph{Full Matrix Representation}. For applications where the original features are the main focus (such as finding driving genes) the full matrix representations $\hat{J}_{k}$ ($d_k \times n$), $k = 1, \ldots, K$ are most useful. These are shown in Figure~\ref{fig:jive:toyjiveoutput}.
	\item \emph{Block Specific Score (BSS)}. For applications where the relationships between subjects are the main focus (such as discrimination between subtypes) large computational gains are available by using the much lower dimensional representations $\hat{\Sigma}^{k}_J \hat{V}^{k}_J$ ($\hat{r}_J \times n$). This results in no loss of information when rotation invariant methods are used. 
	\item \emph{Common Normalized Score (CNS)}. When it is desirable to study the component of joint behavior that is separate from the within block variation (such as evaluating the relationship between data objects), the analysis should focus on a common basis of $\row(\hat{J})$, namely $\hat{V}_J$ ($\hat{r}_J \times n$) from Section~\ref{subsubsec: decomp}.
\end{enumerate}

The relationship between BSS and CNS is analogous to that of the traditional covariance (i.e PLS) and correlation (i.e CCA) modes of analysis. 

Furthermore, different representations provide different views of the loadings. The full matrix representation and BSS naturally obtain the information from the loading matrix $\hat{U}^{k}_J$. CNS gives a different representation of the loadings. Given the common basis of $\row(\hat{J})$, one can perform regression for $\hat{J}_k$ on each score vector in $\hat{V}_J$, from which the standardized coefficient vector can be taken as the CNS loading. By doing this, there is no guarantee of orthogonality between CNS loading vectors. However, the loadings are linked across blocks by their common scores. Therefore, in this CNS case, the standardized regression coefficients are recommended for use instead of the classical loadings.  

The individual variation within blocks can be similarly analyzed resulting in both BSS and CNS analyses for the individual components. When original features are important, the full matrix 
$$\hat{I}_{k} = \hat{U}^{k}_I \hat{\Sigma}^{k}_I \hat{V}^{k}_I, \quad  k = 1, \ldots, K$$
%$$\hat{I}_{Y} = \hat{U}^{Y}_I \hat{\Sigma}^{Y}_I \hat{V}^{Y}_I$$
with dimension $d_k \times n$ are available. Otherwise large computational savings are available from the BSS version $\hat{\Sigma}^{k}_I \hat{V}^{k}_I$ ($\hat{r}_{I_k} \times n$), $k = 1, \ldots, K$. For studying scale free behaviors, use the \emph{Individual Normalized Score (INS)}  $\hat{V}^{k}_I$ ($\hat{r}_{I_k} \times n$). For individual components, the matrix $\hat{U}^{k}_I$ can be taken as loadings for all three representations as the INS matrices cannot be the same. 

\section{Data Analysis}
\label{c:jive-s:datanalaysis}
In this section, we apply Non-iterative JIVE to two real data sets, TCGA breast cancer data set~\citep{cancer2012comprehensive} and Spanish mortality as analyzed in~\citet{marron2014overview}. Detailed analyses are given in Sections~\ref{subsec:TCGA} and~\ref{subsec:mortality} respectively.

\subsection{TCGA Data}
\label{subsec:TCGA}
A prominent goal of modern cancer research, of which TCGA is a major resource, is the combination of biological insights from multiple types of measurements made on common subjects. JIVE is a powerful new tool for gaining such insights. Here we perform our JIVE on gene expression (GE), copy number (CN), reverse phase protein arrays (RPPA) and gene mutation (Mutation) measured on a common set of $616$ breast cancer samples. A most interpretable and insightful analysis is generated from low rank approximations of dimensions $11$ (GE), $6$ (CN), $8$ (RPPA) and $12$ (Mutation) selected in the first step of JIVE. 

\begin{figure}[htb!]
	\centering
	\vspace{4ex}
	\begin{minipage}[b]{0.48\textwidth}
		\centering
		\includegraphics[scale=0.35]{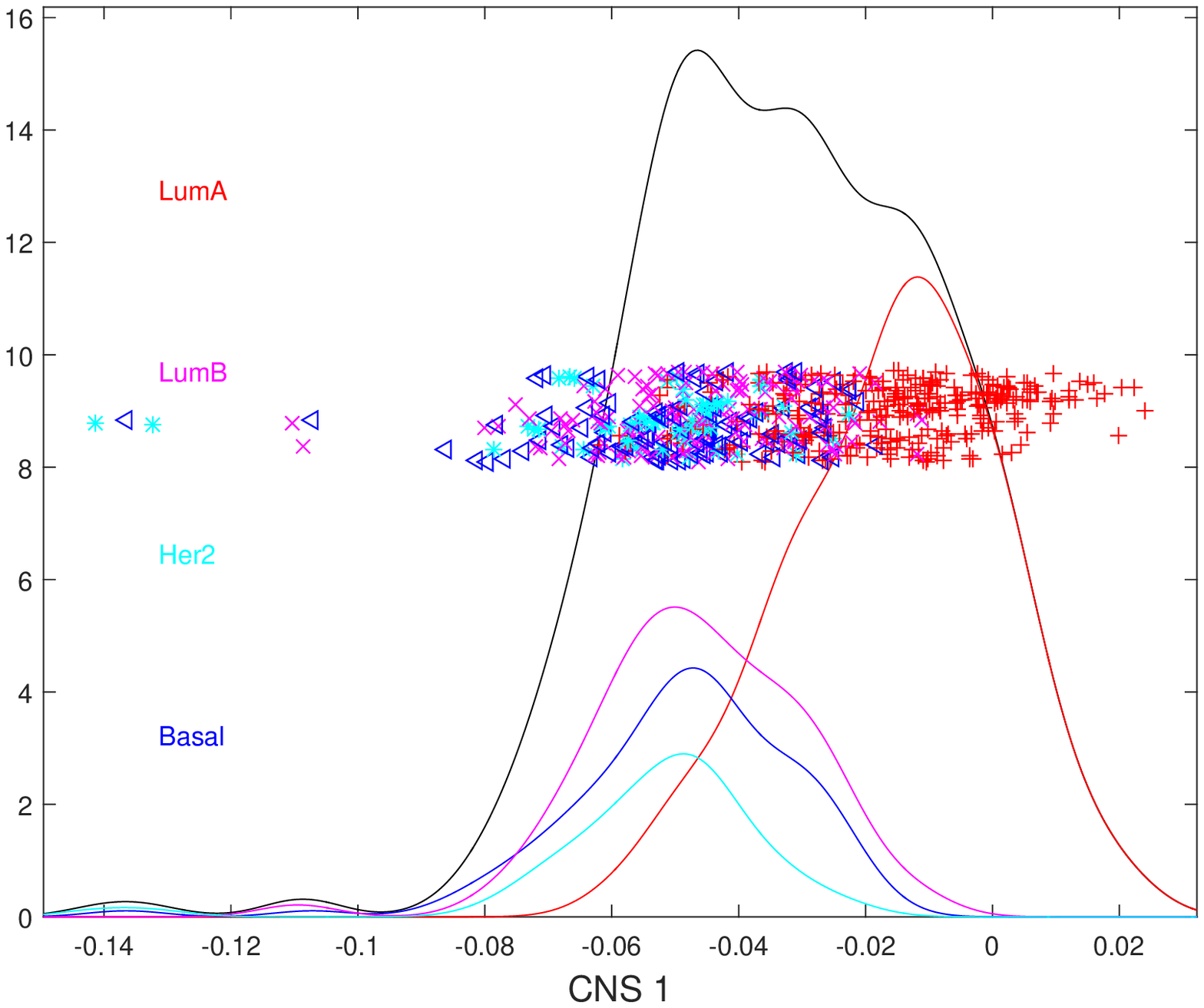}
	\end{minipage}
	\begin{minipage}[b]{0.48\textwidth}
		\centering
		\includegraphics[scale=0.35]{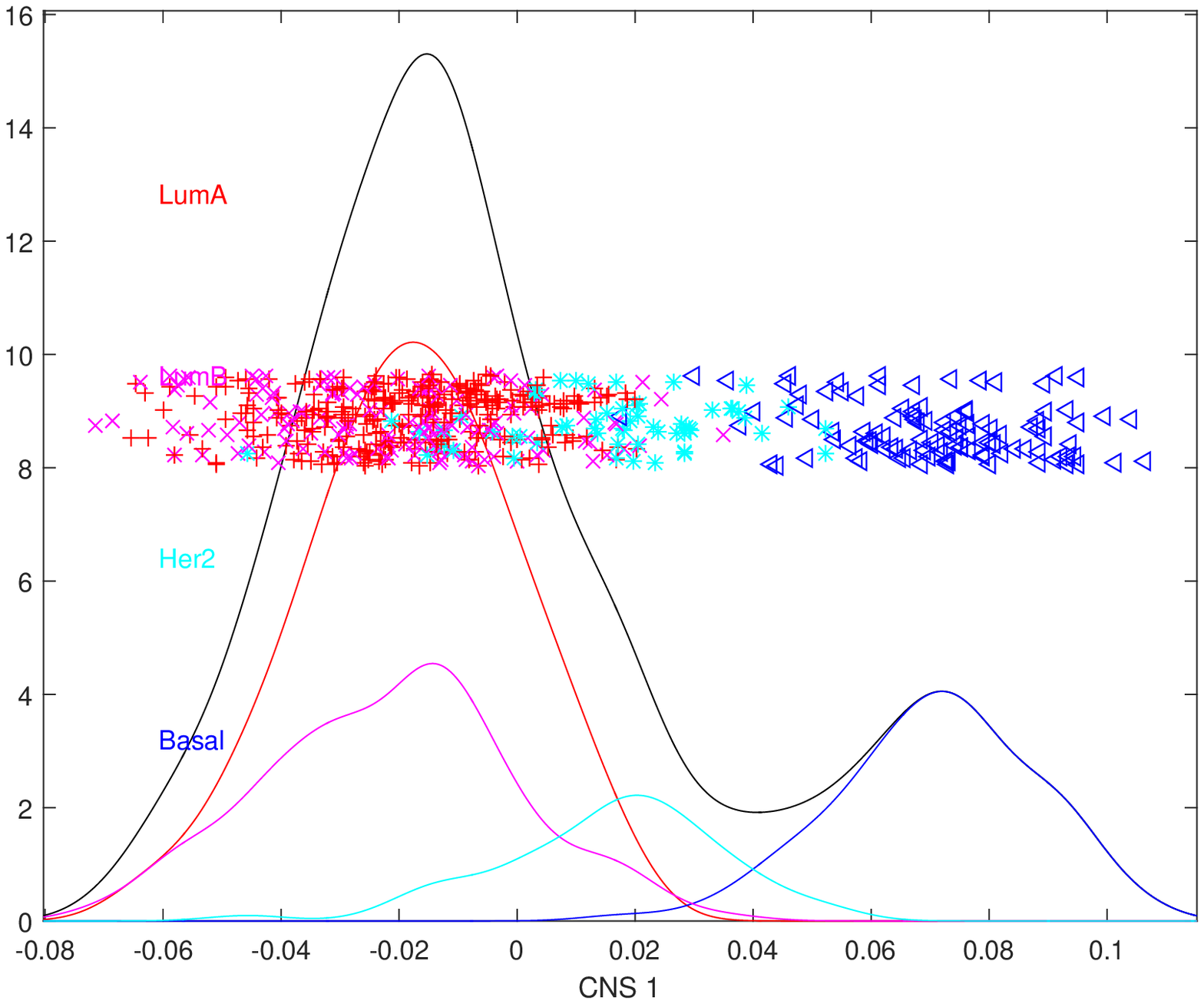}
	\end{minipage}
	\caption{Left: The kernel density estimations of the CNS among GE, CN, RPPA and mutation. The clear separation between Luminal A versus the other subtypes indicates that these four data blocks share a very strong Luminal A property captured in this joint variation component; Right: The CNS from applying JIVE to the individual matrices of GE, CN, and RPPA. The clear separation between kernel density estimations indicates that the individual matrices of GE, CN, RPPA contains a joint variation component explaining the subtype difference between Basal versus the others.}
	\label{fig:tcgamultijive:joint1}
\end{figure}

In the second JIVE step, the one sided 95\% confidence interval suggests selection of two joint components. However, the third step suggests dropping one joint component, because the norm of the projection of the mutation data on that direction, i.e. second CNS, is below the threshold from Step 1. A careful study of all such projections shows that the other data types, i.e. GE, CN and RPPA, do have a common second joint component as discussed at the end of this section. The association between the CNS and genetic subtype differences is visualized in the left panel of Figure~\ref{fig:tcgamultijive:joint1}. For a better understanding of this variation pattern, the dots are a jitter plot of the patients, using colors and symbols to distinguish the subtypes (Blue for Basal-like, cyan for HER2, red for Luminal A and magenta for Luminal B). Each symbol is a data point whose horizontal coordinate is the value and vertical coordinate is the height based on data ordering. The curves are Gaussian kernel density estimates i.e. smoothed histograms, which show the distribution of the subtypes.

The clear separation among density estimates suggest that this joint variation component is strongly connected with the subtype difference between Luminal A versus the other subtypes. To quantify this subtype difference, a test is performed using the CNS of this joint component evaluated by the DiProPerm hypothesis test~\citep{wei2015direction} based on 100 permutations. Strength of the evidence is usually measured by permutation p-values. However, in this context empirical p-values are frequently zero. Thus a more interpretable measure of strength of the evidence is to provide DiProPerm z-score. This is $29.32$ for this CNS. An area under the receiver operating characteristic (ROC) curve (AUC)~\citep{hanley1982meaning} of $0.915$, is also obtained to reflect the classification accuracy. These numbers confirm the strong Luminal A property shared by these four data types. 

A further understanding can be obtained by identifying the feature set of each data type which jointly works with the others in characterizing the Luminal A property. By studying the loading coefficients, important mutation features TP53,TTN and PIK3CA are identified which are well known features from previous studies. Similarly the dominants of GATA3 in RPPA is well known, and is connected with the large GATA3 mutation loading. A less well known result of this analysis is the genes appearing with large GE loadings. Many of these are not dominant in earlier studies, which had focused on subgroup separation, instead of joint behavior. 

As noted in the discussion of Step 2 above, which all four data types have only one significant joint component, the individual components for all of GE, CN and RPPA seem to have $3$--way joint components. This is investigated by a second JIVE analysis of the 3 block concatenation of those individual variation matrices from the initial analysis. The JIVE analysis results in one joint variation component which is displayed in the right panel of Figure~\ref{fig:tcgamultijive:joint1}. This joint variation component clearly shows the differences among Basal, HER2 and Luminal subtypes. In particular, a subtype difference between Basal-like versus the others is quantified using the DiProPerm z-score ($29.82$) and the AUC ($0.998$). Considering the fact that the AUC of the classification between Basal-like versus the others using all the original separate GE features is $0.999$, this single joint component contains almost all the variation information for separating Basal-like from the others. This hierarchical application of JIVE reveals an important joint component that is specific to GE, CN and RPPA but not to Mutation.

\subsection{Spanish Mortality Data}
\label{subsec:mortality}
A quite different data set from the Human Mortality Database is studied here, which consists of both Spanish males and females. For each gender data block, there is a matrix of \emph{mortality}, defined as the number of people who died divided by the total, for a given age group and year. Because mortality varies by several orders of magnitude, the $\log_{10}$ of the mortality is studied here. Each row represents an age group from 0 to 95, and each column represents a year between 1908 and 2002. In order to associate the historical events with the variations of mortality, columns (i.e. mortality as a function of age) are considered as the common set of data objects of each gender block. \cite{marron2014overview} performed analysis on the male block and showed interesting interpretations related to Spanish history. Here we are looking for a deeper analysis which integrates both males and females by exploring joint and individual variation patterns. 

Non-iterative JIVE is applied to the two gender blocks centered by subtracting the mean of each age group. The most interesting JIVE analysis comes from 3 male and 3 female components. The resulting JIVE gives 2 joint components and 1 of each individual component. Since the loading matrices provide important information of the effect of different age groups, BSS analysis together with loading matrices is most informative here. 

\begin{figure}[ht!]
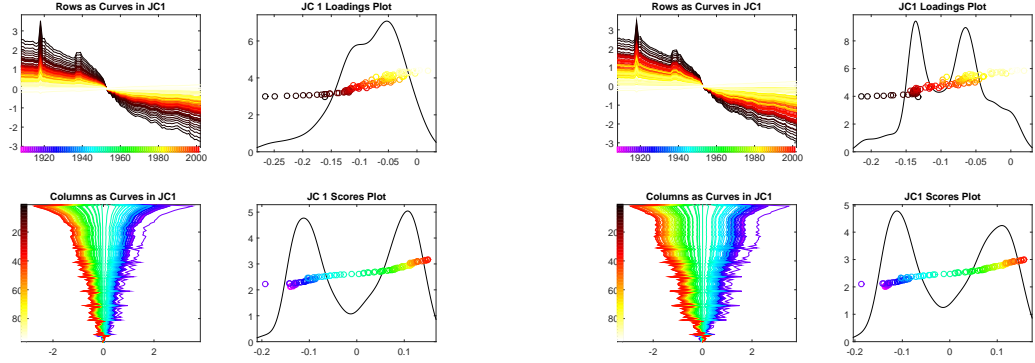

	\begin{minipage}[b]{0.48\textwidth}
		\centering
		\includegraphics[scale=0.35]{JIVEonMaleJointComp1.eps}
	\end{minipage}%%
	\begin{minipage}[b]{0.48\linewidth}
		\centering
		\includegraphics[scale=0.35]{JIVEonFemaleJointComp1.eps}
	\end{minipage}
	\caption{The first BSS joint components of male (left panel) and female (right panel) contain the common modes of variation caused by the overall improvement across different age groups, as can be seen from the scores plots in the right bottom of each panel. The dramatic decrease happened around the 1950s shown in the column projection plot. The degree of decrease degrees varies over age groups.}
	\label{fig:mortalityjoint1}
\end{figure}

Figure~\ref{fig:mortalityjoint1} shows a view of the first joint components for the males (left) and females (right) that is very different from the heat map views used in Section~\ref{c:jive-s:intro-subsec:toy}. While these components are matrices, additional insights come from plotting the rows of the matrices as curves over year (top) and the columns as curves over age (bottom). The curves over year (top) are colored using a heat color scheme, indexing age (black = 0 through red = 40 to yellow = 95 as shown in the vertical color bar on the bottom left). The curves over age (bottom) are colored using a rainbow color scheme (magenta = 1908 through green = 1960 to red = 2002, shown in the horizontal color bar in the top) and use the vertical axis as domain with horizontal axis as range to highlight the fact that these are column vectors. Additional visual cues to the matrix structure are the horizontal rainbow color bar in the top panel, showing that year indexes columns of the data matrix and the vertical heat color bar (bottom) showing that age indexes rows of the component matrix. Because this is a single component, i.e. a rank one approximation of the data, each curve is a multiple of a single eigenvector. The corresponding coefficients are shown on the right. In conventional PCA/SVD terminology, the upper BSS coefficients are called \emph{loadings}, and are in fact the entries of the left eigenvectors (colored using the heat color scale on the bottom). Similarly, the lower coefficients are called \emph{scores} and are the entries of the right eigenvectors, colored using the rainbow bar shown in the top. 

The scores plots together with the rows as curves plots in Figure~\ref{fig:mortalityjoint1} indicate a dramatic improvement in mortality after the 1950s for both males and females. The scores plots are bimodal indicating rapid overall improvement in mortality around the the 1950s. This is also visible in the rows as curves plot. Thus the first mode of joint variation is driven by overall improvement in mortality. In addition to the overall improvement, the rows as curves and scores plots also show the major mortality events, the global flu pandemic of 1918 and the Spanish Civil war in the late 1930s. The loading plots together with the columns as curves plots present the different impacts of this common variation on different age groups for males and females.  The loadings plot for males suggests the improvement in mortality is gradually increasing from older towards younger age groups. In contrast, the female block has a bimodal kernel density estimate of the loadings. This shows that female of child bearing age have received large benefits from improving health care. This effect is similarly visible from comparing the female versus male columns as curves. 

\begin{figure}[ht!]
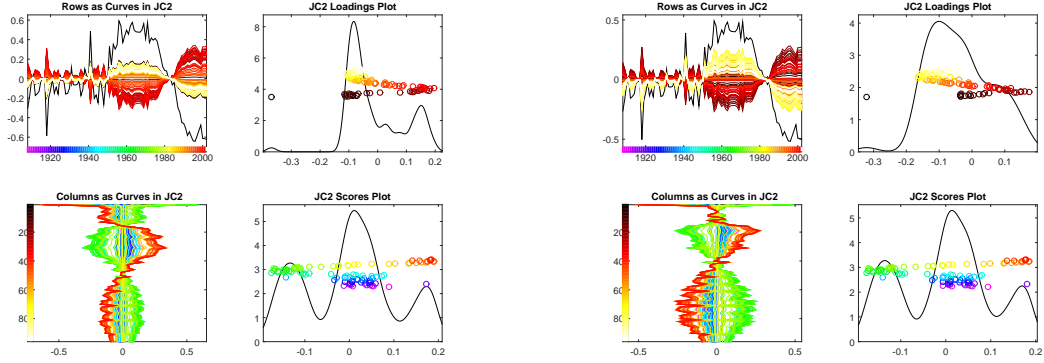
	
	\begin{minipage}[b]{0.48\textwidth}
		\centering
		\includegraphics[scale=0.35]{JIVEonMaleJointComp2.eps}
		\vspace{4ex}
	\end{minipage}%%
	\begin{minipage}[b]{0.48\textwidth}
		\centering
		\includegraphics[scale=0.35]{JIVEonFemaleJointComp2.eps}
		\vspace{4ex}
	\end{minipage}
	\caption{The second joint components of male (left) and female (right) contain the common modes of variation driven by the increase in fatalities caused by automobile penetration and later improvement due to safety improvements. This can be seen from the scores plots in the right bottom. The loadings plots show that this automobile event exerted a significantly stronger impact on the 20-45 males.}
	\label{fig:mortalityjoint2}
\end{figure}

The second BSS components of joint variation within each gender are similarly visualized in Figure~\ref{fig:mortalityjoint2}. This common variation reflects the contrast between the years around 1950 and the years around 1980 which can be told from the curves in the left top and the colors in the right bottom subplots in both male and female panels. In scores plot, the green circles, seen on the left end, represent the years around 1950 when automobile penetration started. And the orange to red circles on the right end correspond to recent years, and much improved car and road safety. The upper left loadings plot of males shows that these automobile events had a stronger influence on the 20-45 males in terms of both larger values and a second peak in the kernel density estimate. Although this contrast can also be seen in the loadings plot of females, it is not as strong as for the male block. The JC2 loadings plots show an interesting outlier, the babies of age zero. We speculate this shows an improvement in post-natal care that coincidently happened around the same time.  

\begin{figure}[ht]
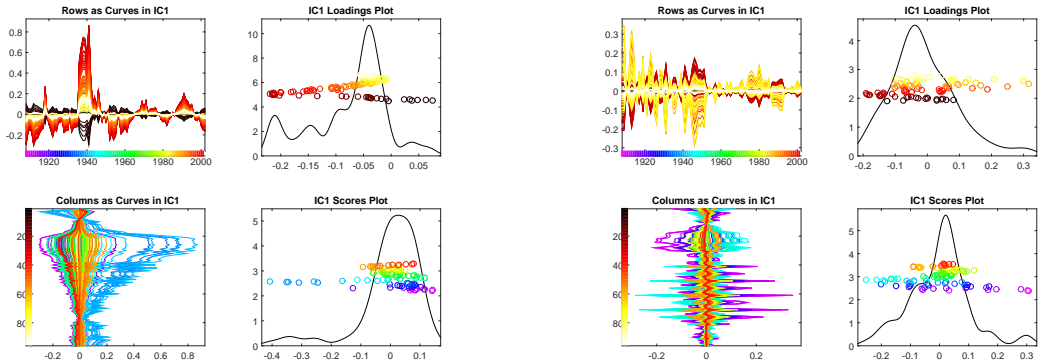
	
	\begin{minipage}[b]{0.48\textwidth}
		\centering
		\includegraphics[scale=0.35]{JIVEonMaleIndividualComp1.eps}
	\end{minipage}%%
	\begin{minipage}[b]{0.48\textwidth}
		\centering
		\includegraphics[scale=0.35]{JIVEonFemaleIndividualComp1.eps}
	\end{minipage}
	\caption{The individual component of male (left) contains the variation driven by Spanish civil war which can be seen from the blue circles on the right end of the right bottom plot. The Spanish civil war mainly affected the young to middle age males.}
	\label{fig:mortalityindiv}
\end{figure}

Another interesting result comes from the studying first individual components (IC1) of males and females, shown in Figure~\ref{fig:mortalityindiv}. In the scores plot of males (left), the blue circles stand out from the rest, corresponding to the years of the Spanish civil war when a significant spike can be seen in the rows as curves plot. Young to middle age groups are affected more than the others which can be found in the loadings plot and columns as curves plot. This year variation pattern, however, cannot be detected in the individual variation component of females. The columns as curves plot on the lower left suggest some type of 5-year age rounding effect, which is seen to occur mostly during the earlier years as indicated both in the rows as curves plot and the colors of the peaks in the columns as curves plot. Note that the plot scales show that the individual female effects are much smaller in magnitude than the male effects.  

\section*{Acknowledgements}
This research was supported in part by the National Science Foundation under
Grant No. 1016441 and 1512945.

\section*{Appendix}
\begin{proof}[ of Theorem~\ref{lemma-exist}]
	Define the row subspaces respectively for each matrix $A_k$ as $\row(A_k) \subseteq \mathbb{R}^n$. For each row subspace, there exists a corresponding projection matrix $P_k$ $(n \times n)$ which is idempotent and symmetric. 
	For non-trivial cases, define a subspace $\row(J) \neq \{\vec{0}\}$ as the intersection of row spaces of $\{\row(A_k), \ k = 1, \ldots, K\}$ i.e.
	$$\row(J) \triangleq \bigcap\limits_{k=1}^{K} \row(A_k).$$
	The projection matrix of subspace $\row(J)$, $P_J$, can thus be represented as $P_J = \prod_{k=1}^K P_k$. Then for each matrix $A_k$, two matrices $J_k$, $I_k$ can be obtained using projection matrix $P_J$ and its orthogonal complement $P_{I_k} \triangleq P_k - P_J$ i.e. $J_k = A_kP_J$ and $I_k = A_kP_{I_k}$. The two matrices satisfy $J_k + I_k = A_k$ and their row subspaces are orthogonal with each other $\row(J) \perp \row(I_k)$, $k = 1, \ldots, K$. 
	
	Moreover, the intersections of row subspaces $\{\row(I_k), \ k = 1, \cdots, K\}$, $\bigcap\limits_{k=1}^{K} \row(I_k)$, has a projection matrix written as 
	\begin{equation*}
	\prod_{k=1}^K P_{I_k} = \prod_{k=1}^K P_{k}(I - P_J) = \prod_{k=1}^K P_{k} \prod_{k=1}^K(I - P_J)  = \boldmath{0} 
	\end{equation*}
	Therefore, we have $\bigcap\limits_{k=1}^{K} \row(I_k) = \{\vec{0}\}$ satisfied and obtain a set of matrices simultaneously satisfying the stated constraints.	
	
	Next we show the sets of matrices $\{J_k \ (d_k \times n), \ k = 1, \ldots, K\}$ and $\{I_k \ (d_k \times n), \ k = 1, \ldots, K\}$ are uniquely defined. Assume the row subspace of matrices $J_k$, $\row(J_k) = \row(J)$, is spanned by a set of bases $\{\vec{v}_1, \ldots, \vec{v}_J\}$ and the row subspaces of $I_k$, $\row(I_k)$, is spanned by a set of bases $\{\vec{w}_1, \ldots, \vec{w}_{Ik}\}$. The row subspace $\row(A_k)$ is thus spanned by their union i.e. $\{\vec{v}_1, \ldots, \vec{v}_J, \vec{w}_1, \ldots, \vec{w}_{Ik}\}$,  since $\row(J_k) = \row(J) \perp \row(A_k)$ for all $k$. Hence, given an arbitrary vector $\vec{v} \in \row(J)$, we always has $\vec{v} \in \row(A_k)$ for all $k$, which indicates 
	$$\row(J_k)= \row(J) \nsubseteq \row(A_k), \quad k = 1, \ldots, K,$$	
	and therefore
	$$\row(J) \subseteq \bigcap\limits_{k=1}^{K} \row(A_k).$$
	Furthermore, suppose there exist a non-zero vector $\vec{a} \in \bigcap\limits_{k=1}^{K} \row(A_k)$ but $\vec{a} \notin \row(J)$ and $\vec{a} \perp \row(J)$. This vector should have $\vec{a} \in \row(I_k), \ k = 1, \ldots, K$ and thus $\vec{a} \in \bigcap\limits_{k=1}^{K} \row(I_k)$ which contradicts the constraint $\bigcap\limits_{k=1}^{K} \row(I_k) = \{\vec{0}\}$. This implies that the row subspace $\row(J)$ is uniquely defined as
	$$ \row(J) = \bigcap\limits_{k=1}^{K} \row(A_k).$$
	Accordingly, the matrices $J_k$ and $I_k$ for $k = 1, \ldots, K$ are also uniquely defined. Otherwise assume there have another set of matrices $A_k = \tilde{J}_k + \tilde{A}_k$ and $P_J$ is the projection matrices of $\row(J)$, we have $J_k = A_kP_J = \tilde{J}_k$.
\end{proof}

\begin{proof}[ of Lemma~\ref{lemma-bound}]
	Let $P_{1}$ and $P_{2}$ be the projection matrices onto the individually perturbed joint row spaces. And let $P$ be the projection matrices onto the common joint row space $J$. Thus, we have
	\begin{align}
	\sin\theta & = \|(I - P_{1})P_{2} \| \\
	& \leq \|(I - P_{1})(I - P)P_{2}\| + \|(I - P_{1})PP_{2}\| \\
	& \leq \|(I - P_{1})(I - P)\|\|(I - P)P_{2}\| + \|(I - P_{1})P\|\|PP_{2}\|
	\end{align}
	in which $\|(I - P_{1})P\| = \sin\theta_{1}$, $\|(I - P_{1})(I - P)\| = \cos\theta_{1}$, $\|(I - P_{2})P\| = \sin\theta_{2}$ and $\|(I - P_{2})(I - P)\| = \cos\theta_{2}$.
	Therefore,
	\begin{equation*}
	\sin\phi \leq \cos\theta_{1}\sin\theta_{2} + \sin\theta_{1}\cos\theta_{2} = \sin(\theta_{1} + \theta_{2}).
	\end{equation*}
\end{proof}

\begin{proof}[ of Lemma~\ref{lemma-bound-multi}]	
	Denote the spanning basis for the estimates of each signal score spaces $\row(\tilde{A}_k)$ as $\{\tilde{V}_k, k = 1, \cdots, K\}$ and $M$ as the vertical concatenation of right singular vector matrices $\{\tilde{V}_k^T, k = 1, \cdots, K\}$ (denoted as $M$) for SVD.
	$$
	M \triangleq \begin{bmatrix}
	\tilde{V}_{1}^{T} \\
	\vdots \\
	\tilde{V}_{K}^{T} \\
	\end{bmatrix} = U_{M}\Sigma_{M}V_{M}^{T}.
	$$
	For each singular value, it can be formulated as a sequential optimization problem i.e
	$$ \sigma_i^2 = \textnormal{max} \|MQ\|^2_{F} = \textnormal{max} \sum_{k=1}^K\|\tilde{V}_{1}^{T}Q\|^2_{F},$$
	in which $Q$ is a rank $1$ projection matrix that is orthogonal to the previous $i-1$ optima i.e. $Q_{1}, \ldots, Q_{i-1}$. For the one that maximizing the Frobenius norm of $M$ projected onto it i.e. $\sigma_i$, we denote as $Q_i$. 
	
	For an arbitary component in the theoretical joint score subspace $\row(J)$, write its projection matrix as $P_J^{(1)}$. The Frobenius norm of $M$ projected onto $P_J^{(1)}$ is 
	\begin{equation}
	\|MP_J\|^2_{F}   =  \begin{bmatrix}
	\tilde{V}_{1}^{T}P_J^{(1)} \\
	\vdots \\
	\tilde{V}_{K}^{T}P_J^{(1)} \\
	\end{bmatrix}_F^2
	\geq  \begin{bmatrix}
	\cos{\theta_1} \\
	\vdots \\
	\cos{\theta_K} \\
	\end{bmatrix}_F^2
	= \sum_{k=1}^{K} \cos^2\theta_{k}
	\end{equation}
	
	Considering the mechanism of SVD,  $\sigma_1^2$ is the maximal norm obtained from the optimal projection matrix $Q_1 \subseteq \bigcup_{k=1}^{K}\row(\tilde{A}_k) \subseteq \mathbb{R}^n$. If all $\tilde{A}_k$ contain all components obtained by noise perturbation of the common row space $\row(J)$, then we have 
	$$\sigma_1^2 \geq \|MP_J\|^2_{F} \geq \sum_{k=1}^{K} \cos^2\theta_{k}$$
	to be considered as a component of joint score subspace. 
	
	This argument can be applied sequentially. For the $Q_2 \in Q_1^{\perp} \cap \{\bigcup_{k=1}^{K}\row(\tilde{A}_k)\}$, there exist a non-empty joint subspace ($\subseteq \row(J)$) that all $ Q_1^{\perp} \cap \row(\tilde{A}_k)$ contain perturbed directions of a joint component other than the one above. Therefore this joint component with projection matrix $P_J^{(2)}$ should have
	$$\sigma_2^2 \geq \|MP_J^{(2)}\|^2_{F} \geq \sum_{k=1}^{K} \cos^2\theta_{k}.$$
	This will continue at least $r_J$ steps and the singular values corresponding to the joint components satisfy the inequality above. 
\end{proof}

\bibliographystyle{plainnat}
\bibliography{reference}

\end{document}